\documentclass[a4paper,11pt]{article}
\pdfoutput=1 

\usepackage{jcappub} 

\usepackage[T1]{fontenc} 
\usepackage[dvipsnames]{xcolor}
\definecolor{PColour}{rgb}{0.1,0.7,0.1}
\definecolor{IColour}{rgb}{0.65,0.,0.5}
\definecolor{RColour}{rgb}{0.1,0.2,0.9}

\title{\boldmath Concerns about the reliability of publicly available SNe~Ia~data}

\author{Mohamed Rameez}

\affiliation{Niels Bohr Institute, Blegdamsvej 17, 2100 Copenhagen, Denmark}

\emailAdd{Mohamed.Rameez@nbi.ku.dk}

\abstract{
I highlight several concerns regarding the consistency of Type~Ia supernova data in the publicly available Pantheon and JLA compilations. The measured  heliocentric redshifts ($z_\mathrm{hel}$) of $\sim 150$ SNe~Ia as reported in the Pantheon catalogue are significantly discrepant from those in JLA --- with 58 having differences amounting to between 5 and 137~times the quoted measurement uncertainty. The discrepancy seems to have been introduced in the process of rectifying a previously reported issue. The Pantheon catalogue until very recently had the redshifts of all SNe~Ia up to $z\sim0.3$ modified under the guise of `peculiar velocity corrections' --- although there is no information on peculiar velocities at such high redshifts. While this has reportedly been rectified on Github by removing peculiar velocity corrections for $z>0.08$, the impact of this on the published cosmological analysis of the Pantheon catalogue is not stated. In JLA, the effect of these 'corrections' is to significantly bias the inferred value of $\Omega_\Lambda$ towards higher values, while the equivalent effect on Pantheon cannot be ascertained due to the unavailability of the individual components of the covariance matrix in the public domain. I provide Jupyter notebooks in order to allow the reader to ascertain the veracity of these assertions.}

\begin{document}
\maketitle
\flushbottom

\section{Inconsistent values of $z_\mathrm{hel}$}
\label{sec:inconzhel}

\begin{figure*}
\centering
\includegraphics[width=\columnwidth]{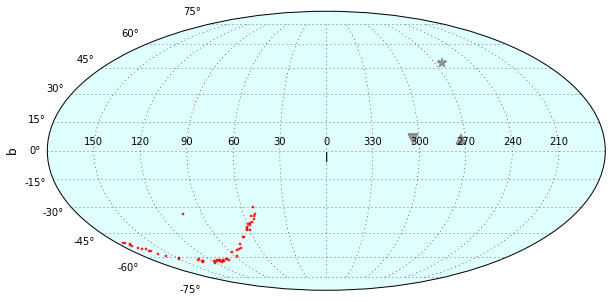} 
\caption{The directions of the 58 SNe documented in Tables \ref{tab:zhel1} and \ref{tab:zhel2}. The directions of the CMB dipole (star), the SMAC bulk flow\cite{Hudson:2004et} (triangle), and the 2M++ bulk flow\cite{Carrick:2015xza} (inverted triangle) are also shown. All the SNe with shifted redshifts are in the hemisphere opposite to the CMB dipole and bulk flow directions.} 
\label{fig:moneyplot}
\end{figure*}

The measured redshifts of SNe~Ia in the heliocentric frame, $z_\mathrm{hel}$ are essential for drawing cosmological inferences, e.g. whether the Hubble expansion rate is accelerating under the influence of dark energy. However for 58 SNe~Ia that are common between the JLA~\citep{Betoule:2014frx} and the Pantheon~\citep{Scolnic:2017caz} compilations, the quoted values are different by $5-137$ times the quoted uncertainty in the redshift measurements. The details are summarised in Tables \ref{tab:zhel1} and \ref{tab:zhel2}, while the distribution of these SNe in the sky can be seen in Fig. \ref{fig:moneyplot}. Many more SNe~Ia have smaller shifts in their $z_\mathrm{hel}$ values.  According to \citep{Kessler:2009ys}, the uncertainty on the spectroscopic redshift measurement of the host galaxy is $1-2 \times10^{-4}$ in SDSS DR4, and 0.0005 for redshifts measured by the authors themselves. (Note that the the $\sigma_{z_\mathrm{spec}}$ arising from peculiar velocities mentioned in \citep{Kessler:2009ys} is the expected dispersion w.r.t theoretical predictions and not the measurement uncertainties.) The quoted redshifts cannot have changed unless if they have been remeasured, a process that has not been documented in \citep{Scolnic:2017caz}. The JLA $z_\mathrm{hel}$ values are in exact agreement with public sources such as VizieR, while the Pantheon $z_\mathrm{hel}$ values are not verifiable independently.

These shifts seem to have been introduced on Nov 27 2018, when new files were uploaded to purportedly rectify previously reported errors\footnote{\href{https://github.com/dscolnic/Pantheon/issues/2}{https://github.com/dscolnic/Pantheon/issues/2}} (section \ref{sec:inconzcmb}).

\subsection{Implications for cosmology}

In \cite{Colin:2018ghy}, my collaborators and I had reported that a dipole in the deceleration parameter aligned with the CMB dipole cannot be statistically distinguished from an isotropic acceleration, due to the directional and redshift distribution of Pantheon SNe~Ia. The former is a natural consequence of an observer located inside a bulk flow, and may be expected given that the bulk flow of the local Universe has not been demonstrated to converge to the CMB rest frame, even up to a depth of 200~$h^{-1}$~Mpc \cite{Carrick:2015xza}.

Out of the 1048 Pantheon SNe~Ia, 890 (including all SDSS SNe~Ia from Tables \ref{tab:zhel1} and \ref{tab:zhel2}) are in the hemisphere \emph{opposite} to the direction of the 2M++ bulk flow\cite{Carrick:2015xza}. Thus a significant change in the redshifts of these SNe~Ia have implications for the dipole and monopole of the deceleration parameter. After the new, shifted redshifts were introduced on Nov 27 2018, works such as \cite{Soltis:2019ryf} and \cite{Zhao:2019azy} relying on these data have appeared, the results of which need to be viewed with significant skepticism.

\begin{centering}
\begin{table*}
\caption{\textcolor{blue}{SNe~Ia with different $z_\mathrm{hel}$ in JLA and Pantheon} : While the JLA and Pantheon names of the SNe~Ia differ by survey specific prefixes, the fact that they are the same can be verified by checking their Right Ascension and Declination coordinates. The JLA redshifts are taken from \href{https://github.com/cmbant/CosmoMC/blob/master/data/jla\_lcparams.txt}{https://github.com/cmbant/CosmoMC/blob/master/data/jla\_lcparams.txt} while the Pantheon redshifts are from \href{https://github.com/dscolnic/Pantheon/blob/master/lcparam\_full\_long\_zhel.txt}{https://github.com/dscolnic/Pantheon/blob/master/lcparam\_full\_long\_zhel.txt}. The shifts in the redshift are computed assuming the conservative value of $\sigma_z = 0.0005$ \citep{Kessler:2009ys}.}
\label{tab:zhel1}
\begin{tabular} 
{| c | c | c | c | c | c |}
\hline
Name in JLA  & JLA $z_\mathrm{hel}$ &  Name in Pantheon   &  Pantheon $z_\mathrm{hel}$   &   $z_\mathrm{diff}$   &   shift  \\ 
\hline\hline
SDSS12881 &  0.233  &  12881 &  0.237838  &  0.004838  &  9.676 $\sigma$ \\
SDSS12927 &  0.175  &  12927 &  0.189638  &  0.014638  &  29.276 $\sigma$ \\
SDSS13044 &  0.121  &  13044 &  0.125735  &  0.004735  &  9.47 $\sigma$ \\
SDSS13136 &  0.366  &  13136 &  0.371627  &  0.005627  &  11.254 $\sigma$ \\
SDSS13152 &  0.207  &  13152 &  0.203311  &  0.003689  &  7.378 $\sigma$ \\
SDSS13305 &  0.201  &  13305 &  0.214557  &  0.013557  &  27.114 $\sigma$ \\
SDSS13727 &  0.221  &  13727 &  0.226402  &  0.005402  &  10.804 $\sigma$ \\
SDSS13796 &  0.145  &  13796 &  0.148518  &  0.003518  &  7.036 $\sigma$ \\
SDSS14261 &  0.281  &  14261 &  0.285517  &  0.004517  &  9.034 $\sigma$ \\
SDSS14331 &  0.214  &  14331 &  0.220905  &  0.006905  &  13.81 $\sigma$ \\
SDSS14397 &  0.371  &  14397 &  0.386084  &  0.015084  &  30.168 $\sigma$ \\
SDSS14437 &  0.144  &  14437 &  0.149098  &  0.005098  &  10.196 $\sigma$ \\
SDSS14481 &  0.255  &  14481 &  0.243249  &  0.011751  &  23.502 $\sigma$ \\
SDSS15057 &  0.299  &  15057 &  0.246586  &  0.052414  &  104.828 $\sigma$ \\
SDSS15203 &  0.216  &  15203 &  0.204218  &  0.011782  &  23.564 $\sigma$ \\
SDSS15287 &  0.274  &  15287 &  0.237419  &  0.036581  &  73.162 $\sigma$ \\
SDSS15301 &  0.248  &  15301 &  0.17963  &  0.06837  &  136.74 $\sigma$ \\
SDSS15365 &  0.178  &  15365 &  0.187733  &  0.009733  &  19.466 $\sigma$ \\
SDSS15383 &  0.312  &  15383 &  0.315791  &  0.003791  &  7.582 $\sigma$ \\
SDSS15440 &  0.253  &  15440 &  0.262051  &  0.009051  &  18.102 $\sigma$ \\
SDSS15461 &  0.18  &  15461 &  0.185954  &  0.005954  &  11.908 $\sigma$ \\
SDSS15704 &  0.365  &  15704 &  0.370275  &  0.005275  &  10.55 $\sigma$ \\
SDSS15868 &  0.242  &  15868 &  0.250516  &  0.008516  &  17.032 $\sigma$ \\
SDSS15872 &  0.203  &  15872 &  0.20629  &  0.00329  &  6.58 $\sigma$ \\
SDSS15897 &  0.17  &  15897 &  0.174692  &  0.004692  &  9.384 $\sigma$ \\
SDSS15901 &  0.199  &  15901 &  0.204563  &  0.005563  &  11.126 $\sigma$ \\
SDSS16072 &  0.277  &  16072 &  0.285523  &  0.008523  &  17.046 $\sigma$ \\
SDSS16073 &  0.146  &  16073 &  0.154541  &  0.008541  &  17.082 $\sigma$ \\
SDSS16116 &  0.15  &  16116 &  0.156305  &  0.006305  &  12.61 $\sigma$ \\
SDSS16185 &  0.097  &  16185 &  0.101255  &  0.004255  &  8.51 $\sigma$ \\
SDSS16206 &  0.152  &  16206 &  0.15954  &  0.00754  &  15.08 $\sigma$ \\
SDSS16232 &  0.367  &  16232 &  0.37532  &  0.00832  &  16.64 $\sigma$ \\
SDSS17220 &  0.172  &  17220 &  0.178821  &  0.006821  &  13.642 $\sigma$ \\
 \hline
\end{tabular}
\end{table*}
\end{centering}

\begin{centering}
\begin{table*}
\caption{\textcolor{blue}{Table \ref{tab:zhel1} (Continued)}}
\label{tab:zhel2}
\begin{tabular} 
{| c | c | c | c | c | c |}
\hline
Name in JLA  & JLA $z_\mathrm{hel}$ &  Name in Pantheon  &  Pantheon $z_\mathrm{hel}$   &   $z_\mathrm{diff}$   &   shift  \\ 
\hline\hline
SDSS17552 &  0.25  &  17552 &  0.253014  &  0.003014  &  6.028 $\sigma$ \\
SDSS17809 &  0.282  &  17809 &  0.288624  &  0.006624  &  13.248 $\sigma$ \\
SDSS18325 &  0.255  &  18325 &  0.258369  &  0.003369  &  6.738 $\sigma$ \\
SDSS18602 &  0.135  &  18602 &  0.138175  &  0.003175  &  6.35 $\sigma$ \\
SDSS18617 &  0.322  &  18617 &  0.326919  &  0.004919  &  9.838 $\sigma$ \\
SDSS18721 &  0.393  &  18721 &  0.402456  &  0.009456  &  18.912 $\sigma$ \\
SDSS18740 &  0.157  &  18740 &  0.154249  &  0.002751  &  5.502 $\sigma$ \\
SDSS18787 &  0.193  &  18787 &  0.190054  &  0.002946  &  5.892 $\sigma$ \\
SDSS18804 &  0.192  &  18804 &  0.198237  &  0.006237  &  12.474 $\sigma$ \\
SDSS18940 &  0.22  &  18940 &  0.212127  &  0.007873  &  15.746 $\sigma$ \\
SDSS19002 &  0.268  &  19002 &  0.27081  &  0.00281  &  5.62 $\sigma$ \\
SDSS19027 &  0.295  &  19027 &  0.2923  &  0.0027  &  5.4 $\sigma$ \\
SDSS19341 &  0.228  &  19341 &  0.236507  &  0.008507  &  17.014 $\sigma$ \\
SDSS19632 &  0.308  &  19632 &  0.314512  &  0.006512  &  13.024 $\sigma$ \\
SDSS19818 &  0.293  &  19818 &  0.304775  &  0.011775  &  23.55 $\sigma$ \\
SDSS19953 &  0.119  &  19953 &  0.123087  &  0.004087  &  8.174 $\sigma$ \\
SDSS19990 &  0.246  &  19990 &  0.24967  &  0.00367  &  7.34 $\sigma$ \\
SDSS20040 &  0.285  &  20040 &  0.287713  &  0.002713  &  5.426 $\sigma$ \\
SDSS20048 &  0.182  &  20048 &  0.185096  &  0.003096  &  6.192 $\sigma$ \\
SDSS20084 &  0.131  &  20084 &  0.139557  &  0.008557  &  17.114 $\sigma$ \\
SDSS20227 &  0.284  &  20227 &  0.276958  &  0.007042  &  14.084 $\sigma$ \\
SDSS20364 &  0.215  &  20364 &  0.218249  &  0.003249  &  6.498 $\sigma$ \\
SDSS21062 &  0.147  &  21062 &  0.13848  &  0.00852  &  17.04 $\sigma$ \\
sn1997dg &  0.0308  &  1997dg &  0.03396  &  0.00316  &  \\
sn2006oa &  0.06255  &  2006oa &  0.059931  &  0.002619  &  \\
 \hline\hline
\end{tabular}
\end{table*}
\end{centering}

\section{Inconsistent, incomplete, and wrong `peculiar velocity corrections'}
\label{sec:inconzcmb}

The catalogues also provide $z_\mathrm{cmb}$, the redshift of each SNe~Ia in the CMB rest frame as inferred from a model of the local peculiar (non-Hubble) velocity field.\footnote{Note that the sortable table at \href{https://archive.stsci.edu/prepds/ps1cosmo/scolnic\_datatable.html}{https://archive.stsci.edu/prepds/ps1cosmo/scolnic\_datatable.html} erroneously reports the \emph{same} values for both $z_\mathrm{CMB}$ and $z_\mathrm{hel}$!} The inconsistencies in the peculiar velocity corrections of JLA have been noted earlier \citep{Colin:2018ghy}. These are:

\begin{itemize}

\item That despite purportedly relying on the flow model of \cite{Hudson:2004et}, the corrections are made beyond the extent of this survey ($z \sim 0.04$), and abruptly fall to zero further beyond, despite \cite{Hudson:2004et} reporting a residual bulk velocity of $687 \pm 203$~km s$^{-1}$, a value which is $>4 $ times larger than the uncorrelated velocity dispersion of $c\sigma_z = 150$ km s$^{-1}$ allowed for in the JLA error budget for cosmological fits. 

\item That SDSS2308 has the same $z_\mathrm{cmb}$ and $z_\mathrm{hel}$, despite being at redshift of 0.14.

\end{itemize}

Significantly more egregious errors are seen in the first version of the Pantheon compilation on Github~\cite{github} wherein peculiar velocity corrections were used to modify the redshifts of SNe~Ia all the way up to $z \sim 0.3$, although no survey of the Universe has gone to such depths, and the information required to make such corrections is simply unavailable. This is now stated on Github~\cite{github} to have been fixed by not making any peculiar velocity corrections for $z>0.08$, but the impact of this major change on the cosmological analysis of this dataset which found >5$\sigma$ evidence for cosmic acceleration\citep{Scolnic:2017caz}  remains undocumented.

\subsection{The `corrections' induce a positive bias on inferring $\Omega_\Lambda$}

Repeating the principled statistical analysis of \cite{Nielsen:2015pga} on the JLA catalogue with $z_\mathrm{hel}$ instead of $z_\mathrm{cmb}$ results in $\Omega_\mathrm{M} = 0.270$ and $\Omega_\Lambda = $ 0.429, which provides only $\sim 1.8\sigma$ evidence for acceleration. Subtracting out the bias corrections to $m_B$ also results in $\Omega_\mathrm{M} = 0.218$ and $\Omega_\Lambda = 0.339$ which is  $<1.5\sigma$ evidence for acceleration. This is in contradiction to Table 11 of \cite{Betoule:2014frx}. This is even more surprising since neither \cite{Riess:1998cb} nor \cite{Perlmutter:1998np} (the original discovery papers) employed SN by SN peculiar velocity `corrections', choosing instead to employ an uncorrelated velocity dispersion of $c \sigma_z = 200$ and 300 km~s$^{-1}$ respectively. JLA and Pantheon continue to employ these dispersions on top of the SN by SN corrections, choosing values of 150 and 250 km~s$^{-1}$ respectively. The justification for the different choices of this dispersion is not given.

Easy to use code to verify these assertions can be found at \footnote{https://github.com/rameez3333/PantheonvsJLAcrosschecks}

Studies such as that above \emph{cannot} be performed on Pantheon, because the covariance matrices corresponding to measurement uncertainties (calibration, model uncertainties, bias, dust and non-Type~Ia)  and externally imposed/cosmological model dependent dispersions (peculiar velocities, $\sigma_z$, lensing) are not provided separately (as they were for JLA) despite requests.

\section{Conclusions}

The reliability of the Pantheon catalogue, especially the reported redshifts (both $z_\mathrm{hel}$ and $z_\mathrm{cmb}$) is questionable. The peculiar velocity corrections applied need to be scrutinised further by the community. The 2.4\% measurement of the local Hubble constant \cite{Riess:2016jrr} also purportedly uses peculiar velocity 'corrections' based on the same flow model \cite{Carrick:2015xza}. Were they also applied initially till z$\sim$0.3 and later corrected? What is the impact?

\end{document}